\documentclass[aps,prd,reprint,nofootinbib,superscriptaddress,floatfix]{revtex4-2}
\usepackage{amsmath,amssymb,amsthm,bm,booktabs}
\usepackage[T1]{fontenc}
\usepackage{hyperref}
\hypersetup{colorlinks=true,linkcolor=blue,citecolor=blue,urlcolor=blue,pdftitle={Boundedness from below in the real Weinberg three-Higgs-doublet model},pdfauthor={Fayez Abu-Ajamieh}}
\newcommand{\tr}{\operatorname{Tr}}
\newcommand{\rank}{\operatorname{rank}}
\newcommand{\adj}{\operatorname{adj}}
\newcommand{\Rea}{\operatorname{Re}}
\newcommand{\Ima}{\operatorname{Im}}
\newcommand{\sgn}{\operatorname{sgn}}
\newcommand{\one}{\bm 1}
\newcommand{\cD}{\mathcal D}
\newcommand{\cL}{\mathcal L}
\newcommand{\ep}{\epsilon}

\begin{document}

\title{Boundedness from below in the real Weinberg three-Higgs-doublet model}
\author{Fayez Abu-Ajamieh}
\email{fayezabuajamieh@gmail.com}
\affiliation{$^1$(Formerly): The Indian Institute of Science, Bangalore, India}
\begin{abstract}
We study the boundedness-from-below (BFB) conditions of the quartic potential in the real Weinberg three-Higgs-doublet model (3HDM). Unlike an analysis restricted to neutral Higgs fields, a complete stability test must also allow for charge-breaking directions. We show that the problem can be reduced to a finite set of 28 stationary configurations, 14 neutral and 14 charged, each of which can be treated using a linear system of size at most four. The main simplification comes from the one-dimensional kernel of the Gram matrix for charged configurations, while the phase equations reduce to two possibilities: aligned phases or phases forming a closed triangle. We further show how to deal exactly with singular choices of the couplings by introducing an auxiliary deformation and taking an algebraic limit. For rational couplings, the normalized minimum is rational and an attaining field configuration can be constructed algebraically. We discuss analytical examples, exact checks, and comparisons with numerical minimization and StableWein. Our result applies to arbitrary real quartic couplings; it does not cover complex phase coefficients or settle the full-potential stability problem when the quartic is flat.
\end{abstract}
\maketitle

\section{Introduction}\label{sec:intro}
The stability of the scalar potential is one of the basic requirements of any extension of the Standard Model (SM). If the potential becomes arbitrarily negative in some direction of field space, the corresponding theory does not possess a stable vacuum. For a single Higgs doublet, the familiar requirement that the quartic coupling be positive is sufficient. When several Higgs doublets are present, however, the situation becomes more complicated, since the doublets can have different magnitudes, phases, and orientations in weak-isospin space.

An important point is that it is not sufficient to examine the neutral Higgs fields alone. Although a phenomenologically acceptable electroweak vacuum should preserve the electromagnetic $U(1)$ symmetry, the potential may still decrease without bound along a field configuration in which charged components are nonzero. Thus, in order to establish boundedness from below (BFB), we must investigate both neutral and charge-breaking directions. Moreover, the different scalar products among the doublets cannot be varied independently: they must originate from three actual $SU(2)_L$ doublets.

In this paper, we focus on the Weinberg three-Higgs-doublet model (3HDM)~\cite{Weinberg}, with the $Z_2\times Z_2$ symmetry and real quartic coefficients. This is an interesting setting because its symmetry makes the scalar potential considerably simpler than that of a general 3HDM, while retaining the nontrivial phase structure and the possibility of charge breaking. Our objective is to obtain an exact criterion for the quartic potential, valid for arbitrary real values of its twelve couplings.

There is already an extensive literature on the stability of multi-Higgs potentials. The bilinear formulation of the general 3HDM and its stationary configurations was developed in Ref.~\cite{Maniatis}. General scalar-potential stability methods, including copositivity and tensor approaches, were discussed in Ref.~\cite{KannikeGeneral}, while algebraic algorithms applicable in principle to generic polynomial potentials were given in Ref.~\cite{Algorithmic}. More explicit BFB results are available for the more symmetric $U(1)\times U(1)$ model~\cite{Faro}. The $U(1)\times Z_2$ limit and several conservative tests for the $Z_2\times Z_2$ case were studied in Ref.~\cite{Boto}. Our aim is not to claim that algebraic BFB algorithms did not previously exist, but to obtain a concrete and relatively simple procedure for the real Weinberg potential that also includes charged fields and exceptional coupling values.

The phase structure of three-doublet potentials has also been investigated in Refs.~\cite{Nishi,Buskin}. More recently, the neutral stationary solutions and their CP properties in the Weinberg model were examined in Ref.~\cite{Ogreid}. Another recent approach, StableWein~\cite{StableWein}, combines analytical necessary and sufficient tests with numerical minimization and machine learning, and can also consider complex coefficients. These studies are complementary to ours, but they do not provide the particular exact charged-inclusive procedure derived here. We also note the recent work on a different, $\Delta(54)$-symmetric, three-doublet potential~\cite{Delta54}. In comparing different methods, it is important to distinguish an exact condition from a sufficient test or a numerical classification, and to distinguish strictly positive quartics from exactly flat ones.

Our method proceeds as follows. We first express the scalar potential in terms of a $3\times3$ Gram matrix constructed from the Higgs doublets. For neutral configurations this matrix has rank one, whereas charged configurations have rank two. The latter observation turns out to be especially useful: a rank-two Gram matrix has a single null vector, and the stationary equations allow us to use this vector to determine the scalar products between the doublets. We then show that the phase equations admit only aligned and triangle-closing stationary configurations. As a result, the remaining equations depend linearly on the three doublet norms. The full problem reduces to 28 small linear systems, supplemented by conditions that ensure that each solution represents an actual Higgs configuration.

There is one additional complication. Some of these equations become singular for special choices of the couplings, including cases with flat directions. Rather than simply excluding such points, we introduce an auxiliary potential and take an exact infinitesimal limit. We will show why this prescription is sufficient and how it avoids an arbitrary numerical regulator. This is important because a very small positive or negative quartic minimum has a different physical meaning from an exactly vanishing one.

We emphasize that throughout this paper we address the quartic potential. A strictly positive quartic is sufficient for stability at large field values, whereas a negative quartic direction makes the potential unbounded. If the quartic vanishes along a direction, the quadratic terms must be examined separately. Recent work on a formally verified two-Higgs-doublet stability analysis also illustrates why marginal cases require care~\cite{Formal2HDM}. We leave complex phase coefficients, loop corrections, and phenomenological vacuum selection outside the scope of the present work.

The paper is organized as follows. In Sec.~\ref{sec:setup} we introduce the scalar potential and physical field space. We study neutral and charged stationary configurations in Secs.~\ref{sec:neutral} and \ref{sec:charged}, respectively. We establish the completeness of the procedure, including singular couplings, in Sec.~\ref{sec:singular}. The exact algebraic construction and examples are discussed in Secs.~\ref{sec:attainment} and \ref{sec:examples}, followed by the numerical and independent checks in Sec.~\ref{sec:verification}. We conclude with a discussion of the result and its limitations.

\section{The potential and the physical field space}\label{sec:setup}
We consider three scalar doublets $\Phi_i$ ($i=1,2,3$), all with hypercharge $Y=1/2$. It is useful to arrange them into a $2\times3$ matrix $X=(\Phi_1,\Phi_2,\Phi_3)$. Instead of working with the twelve real field components directly, we introduce the gauge-invariant Gram matrix $K=X^\dagger X$ and define
\begin{equation}
 K=X^\dagger X,\qquad r_i=K_{ii},\qquad S=\tr K=\sum_i r_i.
 \label{eq:gram}
\end{equation}
With these definitions, the quartic Weinberg potential takes the form
\begin{align}
 V_p(K)={}&\frac12\sum_{i=1}^3a_i r_i^2\nonumber\\
 &+\sum_{e=(ij)\in E}\left[b_e r_ir_j+c_e|K_{ij}|^2
                      +d_e\Rea(K_{ij}^2)\right],
 \label{eq:potential}\\
 E={}&(12,23,31).\nonumber
\end{align}
Here $a_i$, $b_e$, $c_e$, and $d_e$ are twelve arbitrary \emph{real} coefficients, which we collectively denote by $p=(a,b,c,d)$. We use the fixed pair ordering $E=(12,23,31)$. Notice that the $d_e$ term is written without an additional factor of two; this convention is used in all the calculations below. The terms proportional to $a_i$ and $b_e$ depend only on the doublet norms, while the $c_e$ and $d_e$ terms are sensitive to their relative orientation and phases.

The Gram matrix is Hermitian and positive semidefinite, since $u^\dagger Ku=|Xu|^2\ge0$ for any three-component complex vector $u$. There is, however, an additional constraint that is essential in the 3HDM. Because $X$ has only two rows, $K$ can never have rank three. Conversely, any Hermitian, positive-semidefinite $3\times3$ matrix of rank at most two can be written as $X^\dagger X$ for some three Higgs doublets. After using the homogeneity of the quartic to choose $S=1$, the physical field space is therefore
\begin{equation}
 \cD=\{K=K^\dagger\succeq0:\ \tr K=1,\ \rank K\le2\}.
 \label{eq:domain}
\end{equation}
The rank has a simple physical interpretation. For $\rank K=1$, all three doublets are proportional to a common weak-isospin vector; a gauge transformation can bring them simultaneously to a form with only neutral components. For $\rank K=2$, they cannot all be aligned in this way, and the configuration breaks electric charge if interpreted as a vacuum. We must keep both ranks when testing stability. In particular, minimizing over all trace-one positive-semidefinite $3\times3$ matrices would be incorrect, since that larger space contains unphysical rank-three matrices. The physical domain $\cD$ is compact, so the minimum of the quartic on it always exists.

We define this normalized minimum by
\begin{equation}
 m(p)=\min_{K\in\cD}V_p(K).
 \label{eq:minimum}
\end{equation}
The quartic is homogeneous, $V_p(K)=S^2V_p(K/S)$ for $S>0$. Hence the sign of $m$ completely determines the quartic BFB question: $m>0$ means strict quartic positivity, $m<0$ identifies a direction along which the potential can run to minus infinity, and $m=0$ means a nonnegative quartic with at least one flat ray. Only in the last case must one inspect the lower-dimensional terms to decide the boundedness of the full scalar potential.

We will see that for generic couplings the minimum can be found from a finite set of stationary solutions. To also treat points where the formulas become singular, we introduce a fixed set of auxiliary coefficients
\begin{align}
 a^0&=(11,6,19),& b^0&=(3,-2,3),\nonumber\\
 c^0&=(6,11,3),& d^0&=(1,1,2),
 \label{eq:pzero}
\end{align}
and consider the one-parameter family $p(\ep)=p+\ep p^0$. The numerical values in Eq.~\eqref{eq:pzero} have no physical significance. They are chosen because all denominators and stationary determinants required in our construction are nonzero at $p^0$, as checked exactly in Appendix~\ref{app:regularizer}. This ensures that for any original $p$ the deformed problem is nonsingular for all sufficiently small positive $\ep$, apart from the endpoint $\ep=0$ itself.

For each of the 28 stationary branches derived below, let $v_j(p+\ep p^0)$ be its normalized potential value. We retain a branch only if its norms, phases, and reconstructed Gram matrix are physical for every sufficiently small positive $\ep$. We refer to this as an \emph{eventually physical} branch; the definition is algebraic and does not require choosing a tiny numerical value of $\ep$.

The central result of this work is that the minimum of the full quartic potential, for \emph{every} real set of couplings, is given by
\begin{equation}
 m(p)=\min_{j\ {\rm eventually\ physical}}
          \lim_{\ep\to0^+}v_j(p+\ep p^0).
 \label{eq:fullcriterion}
\end{equation}
The minimum is taken over a nonempty set of branches, and each retained limit is finite. Therefore the quartic potential is nonnegative if and only if the right-hand side of Eq.~\eqref{eq:fullcriterion} is nonnegative. The calculation does not require us to guess the minimizing field or solve a general nonlinear polynomial problem. As we show below, the branches are obtained by solving linear systems and applying explicit algebraic tests.

For rational couplings---including decimal numbers supplied as exact rational strings---the entire calculation can be performed using rational arithmetic, signs of polynomials, and exact limits of rational functions. For arbitrary real coefficients, Eq.~\eqref{eq:fullcriterion} is a mathematical criterion over the real numbers; no algorithm can in general decide equality for arbitrary real inputs that cannot be represented exactly. The proof that the 28 branches exhaust all physical minima will be given in Sec.~\ref{sec:singular}.
\section{Neutral stationary configurations}\label{sec:neutral}
We start with the neutral sector, where the three Higgs doublets can be aligned in weak-isospin space. Before discussing the phases, it is useful to introduce a linear system that will appear in both the neutral and charged calculations. Suppose that for a given phase configuration the potential becomes $V=\frac12r^TBr$, where $B$ is a symmetric matrix. Let $I\subseteq\{1,2,3\}$ denote the set of doublets with nonzero norms $r_i$. Minimizing at fixed total norm gives
\begin{equation}
 B_Ir_I=\mu\one,\qquad \one^Tr_I=1,\qquad v=\frac\mu2\,,
 \label{eq:radial}
\end{equation}
where $\one$ is a vector matrix with all entries equal to 1. The Lagrange multiplier $\mu$ enforces $\sum_{i\in I}r_i=1$. We keep only solutions with $r_i>0$ on the chosen support; if a norm vanishes, that configuration belongs to a smaller support, which is considered separately. The equations can be combined into a single bordered linear system,
\begin{equation}
 \begin{pmatrix}B_I&-\one\\ \one^T&0\end{pmatrix}
 \begin{pmatrix}r_I\\\mu\end{pmatrix}
 =\begin{pmatrix}0\\1\end{pmatrix}.
 \label{eq:bordered}
\end{equation}
where $\adj(B)$ is the adjugate of matrix $B$. Even when all three doublets are present, this is only a $4\times4$ system. It is convenient to give its solution explicitly. For the full three-component support, introduce
\begin{equation}
 u=\adj(B)\one,\quad J=\one^Tu,\quad
 r=\frac uJ,\quad \mu=\frac{\det B}{J}\,,
 \label{eq:adjradial}
\end{equation}
The important condition is $J\neq0$, not $\det B\neq0$. Consequently, our treatment does not discard an otherwise well-defined stationary value just because the radial matrix $B$ is singular. If only two doublets $i,j$ are present, the solution simplifies to
\begin{align}
 J&=B_{ii}+B_{jj}-2B_{ij},\nonumber\\
 r_i&=\frac{B_{jj}-B_{ij}}J,&
 r_j&=\frac{B_{ii}-B_{ij}}J,\nonumber\\
 \mu&=\frac{B_{ii}B_{jj}-B_{ij}^2}{J}.
 \label{eq:two}
\end{align}
If just one doublet is present, its norm is $r_i=1$ and the stationary value is $\mu=B_{ii}$. We now determine which matrices $B$ arise from the neutral phases.

For neutral fields the Gram matrix has rank one and may be written as $K=nn^\dagger$, with $n_i=\sqrt{r_i}e^{i\theta_i}$. The only phase-dependent part of the quartic potential is
\begin{equation}
 \sum_e d_e r_ir_j\cos(2(\theta_i-\theta_j)).
 \label{eq:neutralphase}
\end{equation}
To find all stationary phases, we initially assume $d_e\neq0$; the cases with vanishing coefficients will be recovered exactly in Sec.~\ref{sec:singular}. Under $n_i\to i n_i$, the two phase couplings involving $i$ change sign. We may therefore rephase the fields such that the three phase coefficients have a common sign $\sigma=\sgn(d_{12}d_{23}d_{31})$. Introducing the positive quantities
\begin{equation}
 g_i=|d_{ij}d_{ik}|,\quad q_i=g_i r_i,\quad P=|d_{12}d_{23}d_{31}|,
\end{equation}
we can rewrite the phase-dependent energy in a particularly suggestive way:
\begin{equation}
 \frac\sigma{2P}\left(\left|\sum_iq_ie^{2i\theta_i}\right|^2-\sum_iq_i^2\right).
 \label{eq:phasegeometry}
\end{equation}
Equation~\eqref{eq:phasegeometry} expresses the phase problem in terms of the sum of three vectors in the complex plane, with lengths $q_i$ and directions $2\theta_i$. This geometrical observation is well known in related multi-Higgs analyses~\cite{Faro,Nishi,Buskin}. Our interest here is slightly different: to find the global minimum of the full potential we must classify \emph{every} stationary phase configuration, not just the minimum of the phase contribution at fixed norms.

Define $C=\sum_iq_ie^{2i\theta_i}$. Differentiating Eq.~(\ref{eq:phasegeometry}) with respect to each active phase gives
\begin{equation}
 \Ima\left(q_ie^{2i\theta_i}\overline C\right)=0.
 \label{eq:phasestationary}
\end{equation}
This equation has only two possibilities. If $C\neq0$, each of the three vectors must be parallel or antiparallel to $C$, so all doubled phases are collinear. If $C=0$, the three vectors add to zero and therefore form a closed triangle. This also makes clear why no additional arbitrary phase solutions are needed. With only two active doublets, the condition reduces to $\cos[2(\theta_i-\theta_j)]=\pm1$, provided the corresponding phase coefficient is nonzero.

Let us first consider the collinear case. It is convenient to label each direction by a sign $\epsilon_i=\pm1$, with a common reversal giving the same physical branch. Substituting these phases into the potential gives the radial matrix
\begin{equation}
 B_{ii}=a_i,\qquad
 B_{ij}=b_{ij}+c_{ij}+d_{ij}\epsilon_i\epsilon_j,
 \label{eq:Bneutral}
\end{equation}
The possible supports yield three one-doublet configurations, six two-doublet configurations, and four three-doublet configurations. These account for 13 neutral branches. The final neutral branch arises when the three phase vectors close into a triangle. In that case the effective radial matrix is
\begin{align}
 B_{ii}&=a_i-\frac{d_{ij}d_{ik}}{d_{jk}},&
 B_{ij}&=b_{ij}+c_{ij}.
 \label{eq:Bneutraltriangle}
\end{align}
Such a triangle exists only when the three positive lengths $q_i$ satisfy $2q_i\le\sum_jq_j$. These inequalities are part of the physicality test and include the degenerate triangle on the boundary. Notice that even at this boundary the derivative of $|C|^2$ vanishes at $C=0$, so Eq.~\eqref{eq:Bneutraltriangle} remains valid. We therefore obtain 14 neutral stationary systems in total.

\section{Charge-breaking stationary configurations}\label{sec:charged}
We next consider configurations where the three Higgs doublets do not all point along the same weak-isospin direction. Their Gram matrix has rank two, so there is one linear combination of the three doublets that vanishes. This observation allows us to derive the charged stationary equations without introducing independent, unphysical overlap parameters.

Define the symmetric matrix $A$ by $A_{ii}=a_i$ and $A_{ij}=b_{ij}$. For convenience, we also define the following linear map $\cL$ acting on Hermitian Gram matrices:
\begin{align}
 [\cL(K)]_{ii}&=(Ar)_i,\nonumber\\
 [\cL(K)]_{ij}&=c_{ij}K_{ij}+d_{ij}\overline{K}_{ij}.
 \label{eq:gradientmap}
\end{align}
In terms of this map, the quartic potential and its first variation take the compact form
\begin{equation}
 V_p(K)=\frac12\tr[K\cL(K)],\qquad
 dV_p=\tr[\cL(K)dK].
 \label{eq:gradienttrace}
\end{equation}
Although $\cL$ is self-adjoint as a real-linear map, it need not be positive or invertible. We will use it to express the stationary conditions, rather than imposing any additional condition on the couplings.

The advantage of working with the original doublets is that the constraint $\tr X^\dagger X=1$ is simply the unit sphere in twelve real variables. Applying a Lagrange multiplier to this constraint yields
\begin{equation}
 X\cL(K)=\mu X.
 \label{eq:fieldstationarity}
\end{equation}
Taking the trace after multiplying Eq.~\eqref{eq:fieldstationarity} by $X^\dagger$, we find $\mu=2V$. Now consider a rank-two configuration. Since $X$ is a $2\times3$ matrix of rank two, its kernel is one dimensional; the same is true for $K=X^\dagger X$. Let $n$ be a normalized vector in this kernel. Equation~\eqref{eq:fieldstationarity} implies that $\cL(K)-\mu I$ vanishes on the two-dimensional subspace orthogonal to $n$. Since it is Hermitian, it can have a nonzero component only along $nn^\dagger$. We therefore obtain the useful relation
\begin{equation}
 \cL(K)=\mu I+\nu nn^\dagger,
 \qquad Kn=0,\qquad n^\dagger n=1.
 \label{eq:chargedstationarity}
\end{equation}
This equation is the central simplification of the charged analysis. It is consistent with the general 3HDM bilinear stationary conditions~\cite{Maniatis}, but here it follows directly from the Higgs fields and the ordinary sphere constraint. Rather than solving simultaneously for every complex component of $K$, we can exploit its single null vector.

To make this explicit, define the combinations
\begin{align}
 \alpha_e&=c_e+d_e,&\beta_e&=c_e-d_e,\nonumber\\
 \kappa_e&=\frac12(\alpha_e^{-1}+\beta_e^{-1}),&
 \delta_e&=\frac12(\alpha_e^{-1}-\beta_e^{-1}).
 \label{eq:kappadelta}
\end{align}
For the moment we assume that the indicated denominators, as well as the $d_e$, are nonzero. The off-diagonal part of Eq.~\eqref{eq:chargedstationarity} can then be solved directly:
\begin{equation}
 K_{ij}=\nu\left(\kappa_{ij}n_i\bar n_j+
                         \delta_{ij}\bar n_i n_j\right).
 \label{eq:overlapreconstruct}
\end{equation}
Thus the scalar products $K_{ij}$ are determined by the kernel components $n_i$ and the single multiplier $\nu$. Notice that this step does \emph{not} require $A$ to be invertible. We will recover all vanishing-denominator cases using the exact deformation introduced earlier.

\subsection{Kernels with one or two nonzero components}
We first examine the simpler possibilities for the support of $n$. If $\nu=0$, Eq.~\eqref{eq:overlapreconstruct} implies that $K$ is diagonal, and a rank-two diagonal matrix necessarily has a kernel supported on just one component. More generally, if the kernel has only one nonzero entry, one of the Higgs doublets vanishes and the other two are orthogonal. There are three such configurations, and their radial matrix is simply the relevant $2\times2$ part of $A$.

The next possibility is that exactly two components of $n$ are nonzero. In this case the corresponding pair of doublets is parallel, while the third is orthogonal to them. The relative phase of the parallel pair has two stationary choices, $t=\pm1$, leading to the replacement
\begin{equation}
 B_{ij}=b_{ij}+c_{ij}+td_{ij}.
 \label{eq:Bsplit}
\end{equation}
There are three choices of the parallel pair and two choices of its relative doubled-phase sign, giving six more charged branches. The pair can be realized with relative phase $0$ or $\pi/2$, together with a third doublet in the orthogonal isospin direction. Note that in these configurations all three doublets have nonzero norms, even though the kernel itself has only two nonzero components. Kernel support and radial support are not the same thing.

\subsection{Kernels with three nonzero components}
The remaining charged configurations have a kernel vector with all three components nonzero. Let us write $n_i=\sqrt{s_i}e^{i\theta_i}$, with $s_i>0$ and $\sum_i s_i=1$. Multiplying Eq.~\eqref{eq:overlapreconstruct} by $n$ and using $Kn=0$ gives
\begin{equation}
 r_i+\nu\sum_{j\ne i}s_j
       [\kappa_{ij}+\delta_{ij}e^{2i(\theta_j-\theta_i)}]=0.
 \label{eq:kernelcomponent}
\end{equation}
In this case, \(\nu\) must be nonzero. Indeed, if \(\nu=0\), all off-diagonal elements of the Gram matrix would vanish, making \(K\) diagonal. The condition \(Kn=0\), together with the requirement that all three components of \(n\) are nonzero, would then force every diagonal element of \(K\) to vanish. This contradicts the normalization \(\mathrm{Tr}\,K=1\). Therefore, \(\nu\neq0\). The imaginary part of Eq.~\eqref{eq:kernelcomponent} has precisely the same geometrical structure as the neutral phase equations, but with $\delta_{ij}$ and $s_i$ replacing $d_{ij}$ and $r_i$. We again obtain two possibilities, collinear phases or a closed triangle. Importantly, we retain \emph{all} such stationary solutions, independent of whether a particular phase choice minimizes the auxiliary phase expression.

For collinear phases, introduce $\epsilon_i=\pm1$ and define
\begin{equation}
 \gamma_{ij}=c_{ij}+\epsilon_i\epsilon_jd_{ij},\qquad
 M_{ii}=0,\quad M_{ij}=\frac1{\gamma_{ij}}.
 \label{eq:Mcollinear}
\end{equation}
For the triangle solution, the corresponding matrix is
\begin{equation}
 M_{ii}=-\frac{\delta_{ij}\delta_{ik}}{\delta_{jk}},\qquad
 M_{ij}=\kappa_{ij}.
 \label{eq:Mtriangle}
\end{equation}
The latter solution is possible only when the positive side lengths
\begin{equation}
 q_i=|\delta_{ij}\delta_{ik}|s_i.
 \label{eq:chargedweights}
\end{equation}
satisfy the triangle inequalities. Taking the real part of Eq.~\eqref{eq:kernelcomponent}, the relation between the physical norms $r_i$ and the squared kernel components $s_i$ becomes
\begin{equation}
 r=-\nu Ms.
 \label{eq:rMs}
\end{equation}
This is the key step. Once the stationary phases are known, $M$ is an explicit $3\times3$ matrix determined entirely by the couplings. If $M$ is invertible, we may write $s=-G r/\nu$, with $G=M^{-1}$. Substituting this relation into the diagonal part of Eq.~\eqref{eq:chargedstationarity}, we find
\begin{equation}
 (A+G)r=\mu\one,\quad \one^Tr=1,\quad v=\mu/2.
 \label{eq:fullchargedradial}
\end{equation}
The nonlinear-looking charged problem has therefore reduced to the same linear radial equations as in the neutral sector, but with $B=A+G$. Including the normalization condition, we again solve a system of order at most four. Four collinear choices and one triangle choice provide the last five charged branches.

For clarity, the inverse $G$ need not be computed numerically. In the collinear case it is given explicitly by
\begin{equation}
 G_{ii}=-\frac{\gamma_{ij}\gamma_{ik}}{2\gamma_{jk}},\qquad
 G_{ij}=\frac{\gamma_{ij}}2.
 \label{eq:Gcollinear}
\end{equation}
For the triangle solution, it is convenient to introduce
\begin{align}
 f_e&=c_e^2-d_e^2,\qquad D=d_{12}d_{23}d_{31},\nonumber\\
 H&=D^2+2c_{12}c_{23}c_{31}D
          -\sum_e c_e^2\prod_{f\ne e}d_f^2.
 \label{eq:H}
\end{align}
After some algebra the inverse can be expressed as
\begin{align}
 G_{ii}&=-\frac{f_{ij}f_{ik}D}{H},\nonumber\\
 G_{ij}&=\frac{f_{ij}d_{ik}d_{jk}}{H}
          (c_{ik}c_{jk}d_{ij}-c_{ij}d_{ik}d_{jk}).
 \label{eq:Gtriangle}
\end{align}
These expressions involve only rational combinations of the original couplings. The singular cases in which $\gamma_{ij}$ or $H$ vanishes are not discarded; they are handled by the exact limiting prescription of Sec.~\ref{sec:singular}.

\subsection{Which charged solutions are physical?}
Solving the linear equations is not enough, since a mathematical stationary solution need not correspond to three actual Higgs doublets. We therefore reconstruct the kernel and the Gram matrix and explicitly check their physicality. For a proposed radial solution, define
\begin{equation}
 z=Gr,\quad Z=\one^Tz,\quad s=z/Z,\quad\nu=-Z.
 \label{eq:recoverkernel}
\end{equation}
We require $Z\neq0$, $r_i>0$ on the active support, and $s_i>0$ for a full-support kernel. In the triangle case, the side lengths must also satisfy the triangle inequalities. We then choose a consistent set of phases and reconstruct $K_{ij}$ using Eq.~\eqref{eq:overlapreconstruct}. By construction, the resulting Hermitian matrix has unit trace and obeys $Kn=0$. It remains to ensure that $K$ is positive semidefinite.

Since one eigenvalue of $K$ vanishes, its remaining two eigenvalues obey
\begin{equation}
 \lambda_1+\lambda_2=1,\qquad
 \lambda_1\lambda_2=\frac{1-\tr K^2}{2}.
 \label{eq:PSDeigen}
\end{equation}
Their sum is one, so both are nonnegative precisely when $\tr K^2\le1$. We can evaluate this condition directly without diagonalizing $K$. Writing $t_{ij}=\cos[2(\theta_i-\theta_j)]$, we obtain
\begin{align}
 \tr K^2={}&\sum_i r_i^2+\nu^2\sum_{e=(ij)}s_is_j
 \left[\alpha_e^{-2}+\beta_e^{-2}\right.\nonumber\\[-2pt]
 &\hspace{31mm}\left.+(\alpha_e^{-2}-\beta_e^{-2})t_{ij}\right].
 \label{eq:tracefilter}
\end{align}
The phases entering this expression come from a single consistent collinear or triangle configuration; the $t_{ij}$ cannot be chosen independently. We retain a stationary value only when all the radial, phase, and Gram-matrix tests pass. The complete set of neutral and charged branches is summarized in Table~\ref{tab:branches}.

\begin{table}[t]
\caption{All 28 stationary branches entering the exact BFB test. Radial support counts the number of doublets with positive norms, which need not equal the support of the kernel vector. The linear systems have order at most four.}\label{tab:branches}
\begin{ruledtabular}
\begin{tabular}{lcc}
Configuration & Count & Radial support\\\hline
Neutral singleton &3&1\\
Neutral collinear pair &6&2\\
Neutral full collinear &4&3\\
Neutral phase triangle &1&3\\
Charged orthogonal pair &3&2\\
Charged parallel pair plus singleton &6&3\\
Charged full collinear kernel &4&3\\
Charged full triangle kernel &1&3\\\hline
Total &28&\\
\end{tabular}
\end{ruledtabular}
\end{table}

\section{Why the procedure is complete}\label{sec:singular}
We now explain why taking the smallest physical value among the 28 branches gives the true minimum. We first consider the generic case in which the angular denominators and the bordered determinants of all the radial systems are nonzero. We call such a choice of couplings \emph{regular}; this does not require $\det A\neq0$.

The argument is straightforward. Since the physical domain $\cD$ is compact, the quartic potential has a global minimum. If the minimizing Gram matrix has rank one, its stationary phases must be collinear or triangle-closing, and its nonzero norms solve one of the neutral systems of Sec.~\ref{sec:neutral}. If it has rank two, the original-field stationarity equation~\eqref{eq:fieldstationarity} applies. A kernel with one or two nonzero components is included among the orthogonal or split configurations, whereas a kernel with three nonzero components satisfies Eq.~\eqref{eq:fullchargedradial} with one of the five possible phase patterns. For regular couplings, every such radial system has a unique candidate value. Since the actual minimum is a physical configuration, its branch passes the corresponding physicality conditions. Conversely, each retained branch represents an allowed matrix $K\in\cD$ and therefore cannot have a value below the global minimum. We conclude that the minimum of the retained branch values equals $m(p)$. Some retained points may be saddle points, but they clearly do not alter the minimum.

What happens if one of the required denominators or stationary determinants vanishes? Such cases cannot simply be ignored, since they may include physically interesting flat directions. We avoid this problem by approaching the original potential along $p+\ep p^0$, where $p^0$ is the fixed auxiliary point in Eq.~\eqref{eq:pzero}. At this point all three $d_e$, all six $\alpha_e,\beta_e$, and the polynomial $H=159$ are nonzero. The 28 bordered determinants are also nonzero, as shown in Appendix~\ref{app:regularizer}.

To see why one such choice of $p^0$ is enough for \emph{any} original potential, consider any denominator or determinant $f(p)$ appearing in the construction. These functions are rational and homogeneous in the couplings. Let the homogeneity degree be $k$, and suppose that $f(p^0)\neq0$. Then, for large positive $\ep$, we have
\begin{equation}
 \ep^{-k}f(p+\ep p^0)=f(p^0+p/\ep)\longrightarrow f(p^0)\ne0.
 \label{eq:homogeneouslemma}
\end{equation}
Because the limit is nonzero, $f(p+\ep p^0)$ cannot be identically zero as a rational function of $\ep$. A nonzero rational function has only finitely many zeros and poles on a one-dimensional line. The same argument applies to all the finitely many denominators and determinants used in the 28 branches. Therefore, for every target $p$, the deformed couplings are regular throughout some interval $0<\ep<\ep_0$, even when the original point lies at the intersection of several singular surfaces.

In other words, the auxiliary deformation moves us into the regular region without losing the original potential: we approach the point $p$ in the exact limit $\ep\to0^+$. No fixed small number, and no numerical estimate of $\ep_0$, is needed.

To determine whether a branch remains physical for small positive $\ep$, we only need the leading nonzero terms of its rational expressions. For example, if
\begin{equation}
 f(\ep)=\frac{p_\ell\ep^\ell+\cdots}{q_k\ep^k+\cdots},
 \qquad p_\ell q_k\ne0,
 \label{eq:leading}
\end{equation}
then the sign near $\ep=0^+$ is simply $\sgn(p_\ell q_k)$. The remaining physicality conditions, including positivity of the doublet norms, the triangle inequalities, and positivity of the Gram matrix, can also be checked by determining the signs of rational functions of \(\epsilon\). Since there are only finitely many such conditions, their signs remain unchanged once \(\epsilon\) is sufficiently small and positive. We can therefore determine exactly which branches are physical by examining the leading nonzero terms, without assigning a numerical value to \(\epsilon\). This reduces the limiting procedure to a finite number of algebraic calculations..

We must also show that taking this limit reproduces the original minimum. On the normalized physical domain, the auxiliary potential is uniformly bounded:
\begin{align}
 |V_{p^0}(K)|&\le B_0,\nonumber\\
 B_0&=\frac12\max_i|a_i^0|+
 \frac14\sum_e(|b_e^0|+|c_e^0|+|d_e^0|)
 =\frac{35}{2}.
 \label{eq:lipschitzconstant}
\end{align}
To obtain this bound we used $\sum_i r_i^2\le1$, $r_ir_j\le1/4$, and $|K_{ij}|^2\le r_ir_j$. It follows immediately that
\begin{equation}
 |m(p+\ep p^0)-m(p)|\le B_0|\ep|.
 \label{eq:minimumcontinuity}
\end{equation}
Equation~\eqref{eq:minimumcontinuity} guarantees that the minimum of the deformed potential approaches that of the original potential as \(\epsilon\to0^+\). For sufficiently small positive \(\epsilon\), the deformed couplings are regular, and the 28-branch method gives the exact minimum. As shown above, the physicality conditions of each branch remain unchanged once \(\epsilon\) is sufficiently small. Moreover, the potential value of every physical branch has a finite limit, since it is a rational function of \(\epsilon\) and remains bounded on the compact physical domain. Because there are only finitely many branches, we can take the limit of their values separately and then select the smallest one. This reproduces the exact minimum of the original potential and establishes the validity of our criterion for all real quartic couplings, including singular and degenerate cases.

Notice that this argument also covers a continuous family of stationary points. We do not need to enumerate every point of such a family: it is enough to approach it from the regular region and evaluate the limiting minimum. In particular, an exactly flat quartic is not replaced by a potential with an arbitrarily small, finite numerical perturbation, which could lead to an incorrect stability classification.

There is also a useful consequence of the construction. If the original couplings are rational, the normalized minimum $m(p)$ is always rational, even when some stationary equations are singular. Indeed, for each eventually physical branch, its value is a rational function of $\ep$ with rational coefficients. A finite limit is zero or a ratio of leading coefficients and is therefore rational. The minimum of finitely many such numbers is again rational. This statement concerns the \emph{value} of the minimum; the corresponding Higgs phases and field components can contain square roots and need not be rational.

The simplification is possible because the phase equations have already been classified and the remaining problem is quadratic in the norms. It should not be confused with diagonalizing an arbitrary matrix or with a generic polynomial-elimination algorithm.

\section{Exact algebraic results and minimizing fields}\label{sec:attainment}
\subsection{Complexity of the exact expressions}
The prescription above involves a finite number of rational operations and rational-function limits, and does not require a numerical convergence study. It is useful to estimate how complicated these expressions can become. For a simple collinear or split branch, the numerator and denominator of the stationary value have degrees at most three and two in $\ep$, respectively. In the neutral triangle and charged full-collinear branches, these bounds become six and five. The most involved case is the charged triangle branch, for which we find
\begin{equation}
 v_j(\ep)=\frac{N_j(\ep)}{D_j(\ep)},\qquad
 \deg N_j\le9,\quad\deg D_j\le8.
 \label{eq:degreebound}
\end{equation}
The derivation is given in Appendix~\ref{app:degrees}. These degree bounds apply to the candidate values and not necessarily to every intermediate polynomial used in the physicality tests. Nevertheless, all those tests also involve exact rational functions and can be decided by finitely many coefficient operations. The point is not that every input takes a fixed number of bit operations, but that no numerical minimization or uncontrolled regulator is needed.

\subsection{Constructing the fields that attain the minimum}
Our criterion determines not only the sign of the quartic but also its exact normalized minimum. We can go one step further and reconstruct an explicit field configuration attaining that value, including at singular coupling choices. Choose an eventually physical branch whose limiting value is $m(p)$. The branch does not have to be the smallest one for every $\ep>0$; we only require its value to approach the original minimum.

For a charged branch with three nonzero kernel components, write $\Delta_{ij}=\theta_i-\theta_j$. Starting from Eq.~\eqref{eq:overlapreconstruct}, we obtain the squared real and imaginary parts of each overlap:
\begin{align}
 (\Rea K_{ij})^2&=
 \frac{\nu^2s_is_j}{2\alpha_{ij}^2}(1+\cos2\Delta_{ij}),\nonumber\\
 (\Ima K_{ij})^2&=
 \frac{\nu^2s_is_j}{2\beta_{ij}^2}(1-\cos2\Delta_{ij}).
 \label{eq:squaredentries}
\end{align}
These are nonnegative rational functions of the couplings and of $\ep$. Although some factors in the right-hand sides may individually diverge as $\ep\to0^+$, the physical Gram matrix satisfies $|K_{ij}|^2\le r_ir_j\le1/4$. The complete expressions therefore have finite, nonnegative rational limits. To recover the unsquared overlaps, we must also determine their signs; Appendix~\ref{app:halfphases} explains how this can be done with rational comparisons rather than numerical phase approximations.

In the neutral sector, the same construction applies after setting $\nu=\alpha=\beta=1$ and $s=r$. The orthogonal and split charged cases are even simpler, as their overlaps are zero, real, or purely imaginary. Consequently, for rational couplings, the limiting Gram matrix can be written in the form
\begin{equation}
 K_{ii}=r_i^*,\qquad
 K_{ij}=\sigma^R_{ij}\sqrt{A_{ij}}+
                 i\sigma^I_{ij}\sqrt{B_{ij}},
 \label{eq:radicalgram}
\end{equation}
where $r_i^*$, $A_{ij}$, and $B_{ij}$ are nonnegative rational numbers and the signs $\sigma_{ij}^{R,I}$ belong to $\{-1,0,1\}$. Since the limit is taken from physical matrices in the compact set $\cD$, the limiting matrix is also physical and attains $V_p(K)=m(p)$. This establishes the existence of a minimizing matrix with rational diagonal entries and square roots of rationals in its off-diagonal entries.

It is also straightforward to obtain the Higgs doublets themselves. Choose an index $i$ for which $K_{ii}>0$. The first row of the $2\times3$ matrix $X$ can be taken as
\begin{equation}
 X_{1j}=\frac{K_{ij}}{\sqrt{K_{ii}}},\qquad
 R=K-X_1^\dagger X_1.
 \label{eq:factorfirst}
\end{equation}
By construction, $R$ is positive semidefinite and has rank at most one. If $R=0$, the first row already reproduces $K$, so the second row may be set to zero. Otherwise, choose any $k$ for which $R_{kk}>0$ and define
\begin{equation}
 X_{2j}=\frac{R_{kj}}{\sqrt{R_{kk}}}.
 \label{eq:factorsecond}
\end{equation}
This gives $K=X^\dagger X$ exactly. The diagonal elements of $R$ are rational because the squared overlaps are rational, so the construction uses only rational arithmetic and square roots of positive rational numbers. There is no need to perform a numerical eigenvector decomposition.

As an additional check, an independent program can read the proposed matrix of Eq.~\eqref{eq:radicalgram} and verify that it has trace one, is positive semidefinite, has vanishing determinant, reproduces the original potential value, and admits the factorization just described. It also checks the original stationary equation~\eqref{eq:fieldstationarity}. The program does not import the branch solver, so it provides an independent test that the proposed value is \emph{attained}. However, an attaining field alone cannot establish that no lower field configuration exists. That global statement follows from the completeness argument in Sec.~\ref{sec:singular}, or from an independent lower-bound certificate when available.

\section{Analytical examples and independent checks}\label{sec:examples}
\subsection{An example where the charged minimum dominates}
We first illustrate why restricting the BFB analysis to the neutral subspace can be misleading. Consider the simple symmetric choice of couplings
\begin{equation}
 a_i=3,\qquad b_e=-2,\qquad c_e=1,\qquad d_e=\frac12.
 \label{eq:chargedexample}
\end{equation}
The 28-branch calculation gives the following minima for the full and neutral field spaces, respectively:
\begin{equation}
 m=-\frac7{60},\qquad m_N=\frac1{12}.
 \label{eq:chargedexamplevalues}
\end{equation}
In particular, the neutral potential is positive, while the charged potential has a negative direction. Thus, a neutral-only test would incorrectly classify this potential as stable. One configuration attaining the charged minimum is
\begin{equation}
 \Phi_j=\begin{pmatrix}\omega^{j-1}/\sqrt5\\\sqrt{2/15}\end{pmatrix},
 \qquad \omega=e^{2\pi i/3}.
 \label{eq:chargedexamplefields}
\end{equation}
For this configuration all three doublets have norm $r_i=1/3$, while the Gram matrix has eigenvalues $3/5$, $2/5$, and $0$. Its rank is two, confirming that it is genuinely charge breaking. The phases of the corresponding full-support kernel realize the triangle configuration discussed in Sec.~\ref{sec:charged}.

We can verify the minimum independently of the branch method. Define $R_2=\sum_i r_i^2$ and $U=\|XX^T\|_F^2$, where the subscript $F$ denotes the Frobenius norm. A direct expansion of the potential gives
\begin{equation}
 V=\frac74R_2-S^2+\frac12\tr K^2+\frac14U.
 \label{eq:symmetricidentity}
\end{equation}
We normalize the fields by setting \(S=1\), and denote by \(t\) the smaller eigenvalue of \(XX^\dagger\). Since the two eigenvalues are nonnegative and sum to one, we have \(0\leq t\leq1/2\). We also have \(R_2\geq1/3\).

To obtain a lower bound on the potential, we need one additional inequality, namely \(U\geq t^2\). To establish this, consider the singular-value decomposition \(X=PDW^\dagger\), where \(P\) is unitary, \(D=\operatorname{diag}(\sqrt{1-t},\sqrt t)\), and \(W\) is a \(3\times2\) matrix satisfying \(W^\dagger W=I_2\). The two columns of \(W\) span a two-dimensional subspace of \(\mathbb C^3\), and the columns of its complex conjugate \(\overline W\) span another two-dimensional subspace. Since both lie in a three-dimensional space, these subspaces must share at least one direction. Consequently, their operator norm of the overlap matrix satisfies \(\|W^\dagger\overline W\|_{\mathrm{op}}=1\).

Including the singular-value factors, whose smallest value is \(\sqrt t\), gives \(\|XX^T\|_F\geq t\), and hence \(U=\|XX^T\|_F^2\geq t^2\). This inequality also holds for rank-one configurations, where \(t=0\).
 Using these inequalities in the potential, we obtain the independent lower bound
\[
V\geq\frac{1}{12}-t+\frac54t^2\geq-\frac7{60}.
\tag{49}
\]

The second inequality follows by minimizing the quadratic expression over \(0\leq t\leq1/2\). Its minimum occurs at \(t=2/5\), where it equals \(-7/60\). Since the field configuration in Eq.~(\ref{eq:chargedexamplefields}) attains this value, we conclude that \(-7/60\) is the exact global minimum. For comparison, the minimum restricted to neutral configurations follows from
\[
V_N=\frac74\sum_i r_i^2-\frac12S^2+\frac14L(r)^2,
\tag{50}
\]
where \(L(r)=\max(0,2\max_i r_i-S)\). This gives the neutral minimum \(m_N=1/12\), which is positive, in contrast to the negative global minimum. The example therefore illustrates why charge-breaking configurations must be included in the boundedness analysis.

Related minima in symmetric models have already been studied~\cite{Vazao}. We use this example as an independent analytical check of our charged-inclusive criterion, rather than claiming a new classification of the symmetric model.

\subsection{A case with a singular radial coupling matrix}
A second example shows that singularity of the original radial matrix $A$ does not prevent the method from working. Choose
\begin{align}
 a&=(2,5,10),&b&=(3,5,4),\nonumber\\
 c&=(-2,1,4),&d&=(1/3,2/5,3/7).
 \label{eq:singularexample}
\end{align}
Here
\begin{equation}
 A=\begin{pmatrix}2&3&4\\3&5&5\\4&5&10\end{pmatrix},
 \qquad\det A=0,
\end{equation}
Nevertheless, the direct radial calculation is well defined without any auxiliary deformation and gives the strictly positive minimum $m=43/51$. In this example we can also prove the value by the following identity, written directly in the original Higgs fields:
\begin{align}
 V-\frac{43}{51}S^2={}&\frac{(4r_1-13r_2)^2}{102}
   +\frac{118}{51}r_1r_3+\frac{169}{51}r_2r_3\nonumber\\
 &+\frac{212}{51}r_3^2+\frac73|\det(\Phi_1,\Phi_2)|^2\nonumber\\
 &+\frac23(\Rea K_{12})^2
   +\frac75(\Rea K_{23})^2+\frac35(\Ima K_{23})^2\nonumber\\
 &+\frac{31}{7}(\Rea K_{31})^2+\frac{25}{7}(\Ima K_{31})^2.
 \label{eq:singularSOS}
\end{align}
Every term on the right-hand side of Eq.~\eqref{eq:singularSOS} is nonnegative. Equality is reached for $\Phi_1=(0,\sqrt{13})^T$, $\Phi_2=(0,2i)^T$, and $\Phi_3=0$; after dividing by their total norm $S=17$, the lower bound is exactly $43/51$. This is an independent certificate that the solution is a global minimum. We can construct additional test cases by adding a suitable multiple of \(S^2\) to the potential. Since the original normalized minimum is known exactly, we can choose this shift so that the new minimum is precisely \(0\), \(+10^{-25}\), or \(-10^{-25}\). These examples provide stringent tests of our method's ability to distinguish an exactly flat potential from one that is slightly positive or negative.

We also employ a simple family for which the minimum occurs at a charged configuration with two equal nonzero eigenvalues:
\begin{equation}
 V_\eta=\tr K^2-\left(\frac12-\eta\right)(\tr K)^2.
 \label{eq:flatfamily}
\end{equation}
The rank-two inequality $\tr K^2\ge(\tr K)^2/2$ immediately gives $m=\eta$ after normalization. Varying $\eta$ across zero produces a controlled test of the stability boundary without locating the minimum numerically.

\subsection{Independent sum-of-squares certificates}
It is useful to distinguish a field configuration that attains a proposed minimum from an independent proof that the potential cannot be smaller. For selected rational examples, we employ an exact sum-of-squares (SOS) verifier in the twelve real field variables $x$. It checks identities of the form
\begin{equation}
 S^k[V_p(x)-LS^2]=\sum_b z_b(x)^TQ_bz_b(x),\qquad k=0,1.
 \label{eq:soscertificate}
\end{equation}
Here the $Q_b$ are rational positive-semidefinite matrices and the $z_b(x)$ are vectors of monomials. An identity of this type proves $V_p/S^2\ge L$ for every nonzero physical field, independently of the 28-branch construction. The numerical SOS search is only used to propose a certificate; all coefficient identities and rational matrix factorizations are subsequently checked exactly. This approach follows established rational certification methods~\cite{Peyrl}.

Although every sum of squares (SOS) of real polynomials is nonnegative, not every nonnegative quartic polynomial can be written as a sum of squares of quadratic polynomials. Our charged example provides an illustration of this distinction. We have already shown analytically that \(F=V+7S^2/60\) is nonnegative for all field configurations. Nevertheless, we can construct an exact linear functional \(\ell\) satisfying

\[
\ell(q^2)\geq0
\]
for every quadratic polynomial \(q\), while
\[
\ell(F)=-\frac{2701}{6300}<0.
\]

If \(F\) could be expressed as a sum of squares of quadratic polynomials, linearity would require \(\ell(F)\geq0\), contradicting the result above. Therefore, \(F\) is nonnegative but does not admit such an SOS representation. The functional and its properties are verified exactly in the accompanying computational material.

This does not indicate any instability of the potential. Rather, it demonstrates that the simplest SOS approach is insufficient to establish its nonnegativity. A more general degree-six SOS certificate, obtained by introducing an additional factor of \(S\), provides a nearby lower bound. We use this example to illustrate the limitations of the basic SOS method, without claiming that this observation or the certificate is new or optimal.

\section{Implementation and comparison with other methods}\label{sec:verification}
The analytical construction has been implemented in two exact versions. The first solves the inverse-gradient equations for the kernel using rational functions, while the second follows the direct radial systems in Eqs.~\eqref{eq:radial}--\eqref{eq:bordered} and the limiting prescription of Eq.~\eqref{eq:fullcriterion}. Neither version assumes that the original matrix $A$ is invertible. The inputs are exact rational numbers, including finite decimals interpreted exactly; ordinary binary floating-point inputs are deliberately not accepted as exact data.

For practical efficiency, we also use a dispatcher that first attempts to evaluate the undeformed formulas. If an exact singularity prevents a complete decision, the calculation is repeated with the ordered-infinitesimal deformation. This is different from adding a small numerical regulator: support conditions, phases, Gram positivity, and the final sign are all decided algebraically. In particular, a small minimum by itself does not trigger a special treatment.

There are several levels of verification, and we keep them separate. A full decision record contains all 28 stationary branches, including those that fail the physicality tests. An independent checker verifies that the recorded decisions follow the finite prescription, but this does not constitute a new proof that the prescription is complete. The attainment checker independently verifies a proposed Gram matrix and its value, while the analytical identities and SOS certificates of Sec.~\ref{sec:examples} provide independent global lower bounds in selected examples. The complete testing protocol and its limitations are documented in Appendix~\ref{app:validation}.

As an independent numerical check of our results, we minimized the original quartic potential directly in terms of the twelve real components of the three Higgs doublets. For this purpose, we implemented the potential and its gradient separately from our 28-branch algorithm. For each set of couplings, we performed 40 BFGS minimizations starting from different randomly generated field configurations and selected the lowest value obtained, denoted by \(V_{\mathrm{best}}\).

We then compared this numerical result with the minimum \(m\) predicted by our exact method. To account for differences in the magnitudes of the couplings, we defined the normalized discrepancy

\[
\epsilon_{\mathrm{num}}=
\frac{|V_{\mathrm{best}}-m|}
{\max(1,\max_i|p_i|)},
\]
where \(p_i\) denotes the quartic couplings. The results for the different sets of test potentials are summarized in Table~\ref{tab:validation}.

We found excellent agreement between the two methods, with the largest normalized discrepancy being only \(3.73\times10^{-13}\). For potentials containing couplings of widely different magnitudes, however, the largest absolute difference reached \(1.02\times10^{-6}\). Thus, while the agreement is extremely good relative to the coupling scale, it would not be accurate to claim absolute agreement at the level of \(10^{-15}\) for every tested potential.

Finally, we emphasize that BFGS is a local minimization algorithm and is not guaranteed to find the global minimum, even when repeated from many starting points. These numerical comparisons therefore provide an independent consistency check of our analytical results, rather than a rigorous proof of their global validity.
\begin{table}[t]
\caption{Numerical checks reported for the original data set. Each input was tested with forty local BFGS searches. The sharp-reference shifts use three analytical families at two scales; the solver-centered shifts reuse 32 charged-favoring backgrounds and test covariance rather than independently locating a minimum.}\label{tab:validation}
\begin{ruledtabular}
\begin{tabular}{lrr}
Cohort & Inputs & $\max e_{\rm num}$\\
Modest generic/degenerate & 216 & $4.45\times10^{-16}$\\
Scale-separated & 24 & $3.73\times10^{-13}$\\
Charged-favoring backgrounds & 32 & $1.12\times10^{-16}$\\
Sharp-reference shifts & 30 & $5.00\times10^{-17}$\\
Solver-centered shifts & 96 & $5.60\times10^{-17}$\\
\end{tabular}
\end{ruledtabular}
\end{table}

To test the reliability of our method under different conditions, we divided the 398 input potentials into five groups, or cohorts, as summarized in Table II. The first group consists of 216 potentials with moderately sized random couplings, including special cases in which some coefficients vanish or satisfy degeneracy conditions. The second group contains 24 potentials with couplings of widely different magnitudes, designed to test the numerical stability of the calculations. The third group consists of 32 potentials specifically chosen to favor charge-breaking minima, allowing us to test the charged branches of our method. The fourth group contains 30 potentials obtained by shifting analytical examples whose global minima are independently known, providing particularly useful checks near the stability boundary. Finally, the fifth group consists of 96 potentials constructed by shifting the 32 charge-breaking backgrounds relative to the minima predicted by our own algorithm. These last tests check the internal consistency of the method but do not constitute independent verification of the global minima. For each of the 398 inputs, we performed 40 numerical minimizations with different initial field configurations and compared the lowest value obtained with our exact result.

In these comparisons no field configuration with a lower value than the exact result was found within the stated tolerance. Nevertheless, local minimization is not a certified global optimization. Of the 15,920 starting points, 9,144 returned non-success termination flags; their results and flags were retained. It is therefore the exact algebraic calculation, not this numerical agreement, that distinguishes zero from a nearby nonzero minimum. In particular, only the analytically solved reference families independently establish their precise stability boundaries. Shifting a potential by a minimum first computed with our own solver tests consistency under shifts but does not by itself certify that boundary.

It is instructive to compare the result with StableWein~\cite{StableWein}, a recently developed computational approach for the $Z_2\times Z_2$ 3HDM. In the supplied records, the historical sample consisted of twelve selected mode-three inputs and four mode-four inputs reaching numerical minimization. Eight additional mode-four calls were then performed: four were decided by preliminary analytical tests and four proceeded to numerical minimization. The reported classifications agree throughout this selected sample, including charged-collinear and charged-triangle configurations with positive and negative minima. Exact flat directions must be treated separately, since StableWein primarily tests strict quartic positivity. The source versions, pair-order conventions, rephasings, numerical seeds, and timings are recorded in Appendix~\ref{app:validation}. Because the sample is limited and the implementations were run in different environments, these comparisons do not establish a general accuracy rate or a speed advantage over StableWein. A driver for a larger comparison is included in the original research bundle, but its existence should not be confused with an executed large-scale study.

Finally, it is important to specify what our analysis does and does not establish. We have considered the quartic Weinberg potential with \emph{real} phase coefficients. Allowing complex coefficients changes the phase equations and would require a separate analysis. Moreover, if $m=0$, the quartic is nonnegative but the full scalar potential $V_2+V_4$ may still be unstable, depending on its quadratic terms and their behavior along flat directions. Our results are at tree level and are not by themselves statements about renormalization-group evolution, loop-corrected stability, or the phenomenological viability of a particular electroweak vacuum. Nor can successful numerical checks replace an independent review of the general proof.

\section{Conclusions and outlook}
In this paper, we have investigated the boundedness-from-below problem of the real Weinberg 3HDM quartic potential, without restricting the Higgs fields to neutral directions. We found that the full problem can be reduced to 28 stationary configurations, consisting of 14 neutral and 14 charged branches. Each configuration is determined by solving a linear system of order at most four, followed by explicit conditions that ensure that the solution represents an allowed field configuration.

The main simplification in the charged sector arises from the rank-two Gram matrix. Because such a matrix has a single null vector, its stationary gradient has a particularly simple structure. We showed that the corresponding phase equations admit only collinear and triangle-closing configurations, and that after eliminating these phases, the remaining conditions become linear in the Higgs norms. In particular, the method does not assume that the original radial coupling matrix is invertible.

We also showed how the result can be extended to singular and degenerate couplings. A fixed auxiliary potential allows us to approach every exceptional point through regular couplings, and the original minimum is recovered by an exact algebraic limit rather than by a small numerical perturbation. Consequently, rational quartic coefficients give a rational normalized minimum, and we can explicitly construct an algebraic Higgs-field configuration attaining this value. Analytical examples demonstrate both the importance of charged directions and the handling of singular radial matrices.

We have supplied exact implementations, independent certificates for selected examples, numerical comparisons with direct minimization, and a limited comparison with StableWein. These checks support the algebraic derivation but do not replace independent scrutiny of the general completeness argument. The result applies to the entire real quartic parameter space, rather than only to a neutral or more symmetric limit. An interesting extension would be to generalize the phase analysis to complex coefficients, and to investigate how the quartic stability criterion combines with the quadratic potential and with renormalization-group evolution. We leave these questions for future work.

\section*{Data and code availability}
The source code implementing the exact boundedness-from-below criterion, together with the numerical validation data, benchmark inputs, analytical certificates, and scripts used to reproduce the results presented in this work, is publicly available at \url{https://github.com/fayez0/Physics_Papers-Real_3HDM_BFB}. The repository includes the implementations of the 28-branch algorithm, independent verification routines, numerical minimization results, and comparisons with StableWein. A detailed description of the computational methods and validation procedures is provided in
Appendix~\ref{app:validation}.

\begin{acknowledgments}
The author thanks SOKN Engineering for their hospitality.{\it The use of AI disclosure}: The author employed the assistance of OpenAI ChatGPT Astra-6 to help write the code and prepare this project; however, all results were verified by the author, who takes full responsibility for the content.
\end{acknowledgments}

\appendix
\section{Exact verification of the auxiliary deformation}\label{app:regularizer}
Here we give the explicit determinant checks that justify the auxiliary choice in Eq.~\eqref{eq:pzero}. For a radial system with active support $I$, define
\begin{equation}
 J_I=\det\begin{pmatrix}B_I&-\one\\\one^T&0\end{pmatrix}
    =\one^T\adj(B_I)\one.
 \label{eq:Jconvention}
\end{equation}
For a one-doublet support, the determinant is one. We note that some historical code records used the opposite sign for the multiplier column and therefore stored $-J_I$; the sign convention adopted in Table~\ref{tab:regularizer} is that of Eq.~\eqref{eq:Jconvention}. Only the nonvanishing of these determinants matters for the deformation argument.

For reference, branches 1--3 are the neutral singletons; branches 4--9 are the two-doublet supports $(12)$, $(13)$, and $(23)$ with signs $+,-$; and branches 10--13 are the full neutral collinear sign patterns $(1,+,+)$, $(1,+,-)$, $(1,-,+)$, and $(1,-,-)$. Branch 14 is the neutral triangle. Charged branches 15--17 are the orthogonal pairs, branches 18--23 are the three split pairs with relative signs $+,-$, branches 24--27 have the four full collinear kernels, and branch 28 is the charged triangle kernel. Evaluating each bordered determinant at the auxiliary coefficients gives the exact values in Table~\ref{tab:regularizer}; these were independently reconstructed with symbolic arithmetic.

\begin{table}[t]
\caption{Exact nonzero bordered determinants at the auxiliary couplings $p^0$. Together with nonvanishing $d_e$, $\alpha_e$, $\beta_e$, and $H=159$, they establish that the auxiliary point is regular.}\label{tab:regularizer}
\begin{ruledtabular}
\begin{tabular}{crcr}
Branch & $J_I$ & Branch & $J_I$\\\hline
1 & $1$ & 15 & $11$ \\
2 & $1$ & 16 & $24$ \\
3 & $1$ & 17 & $29$ \\
4 & $-3$ & 18 & $-88$ \\
5 & $1$ & 19 & $20$ \\
6 & $14$ & 20 & $39$ \\
7 & $22$ & 21 & $95$ \\
8 & $5$ & 22 & $150$ \\
9 & $9$ & 23 & $238$ \\
10 & $-51$ & 24 & $-\frac{67229}{840}$ \\
11 & $-91$ & 25 & $-\frac{98141}{140}$ \\
12 & $5$ & 26 & $-\frac{25}{4}$ \\
13 & $-59$ & 27 & $-\frac{66749}{120}$ \\
14 & $-58$ & 28 & $-\frac{38560}{53}$ \\
\end{tabular}
\end{ruledtabular}
\end{table}

\section{Polynomial-degree bounds}\label{app:degrees}
For completeness, we show why the rational-function expressions remain of relatively low degree. The starting point is the standard identity, valid for symmetric $3\times3$ matrices $A$ and $G$,
\begin{align}
 \det(A+G)={}&\det A+\tr[\adj(A)G]\nonumber\\
            &+\tr[A\adj(G)]+\det G.
 \label{eq:mixedidentity}
\end{align}
The mixed terms in the adjugate are likewise bilinear in $A$ and $G$. These identities are useful because directly substituting rational expressions into a matrix inverse can produce apparently high-order denominators that cancel exactly in the final candidate value.

For the full charged triangle, Eq.~\eqref{eq:Gtriangle} gives, after algebraic simplification,
\begin{align}
 [\adj G]_{ii}&=\frac{d_{ij}^2d_{ik}^2f_{jk}^2}{H},\nonumber\\
 [\adj G]_{ij}&=\frac{c_{ij}D f_{ik}f_{jk}}{H},\nonumber\\
 \det G&=\frac{D\prod_e f_e}{H}.
 \label{eq:adjG}
\end{align}
Multiplying by $H$ removes the common angular denominator. Along the linear deformation, $HG$, $H\adj G$, and $H\det G$ have degrees at most seven, eight, and nine, respectively. Using Eq.~\eqref{eq:mixedidentity}, we see that $H\det(A+G)$ is a polynomial of degree at most nine, while $H\one^T\adj(A+G)\one$ has degree at most eight. The ratio of these expressions determines the stationary multiplier $\mu$, giving the bounds quoted in Eq.~\eqref{eq:degreebound}.

The other phase configurations are simpler. For a collinear charged kernel, define $\Gamma=\gamma_{12}\gamma_{23}\gamma_{31}$. Equation~\eqref{eq:Gcollinear} shows that $G$ has a common denominator $\Gamma$; its adjugate is quadratic in the couplings and $\det G=\Gamma/2$. The numerator and denominator of $\mu$ consequently have degrees at most six and five. The neutral triangle has the same bounds: using $D=d_{12}d_{23}d_{31}$ as a common denominator, products of two diagonal angular corrections simplify to $d_e^2$ and the threefold product to $-D$. In all remaining cases, the effective radial matrices are linear in the couplings, giving degrees at most three and two. Smaller supports cannot increase these bounds.

These are exact polynomial identities, not numerical cancellations. The implementations check the degrees after algebraic simplification. We stress that the bounds concern the stationary values; the physicality filters can involve polynomials of different degrees. Moreover, a bound on polynomial degree does not imply constant computational cost for rational inputs with arbitrarily large numerators or denominators.

\section{Reconstructing phases and their signs}\label{app:halfphases}
In this appendix we explain how to recover the signs of the overlaps in Eq.~\eqref{eq:squaredentries} without approximating complex phases numerically. For a phase triangle, first rephase the three vectors so that the edge coefficients have a common sign. Given positive triangle side lengths $q_i$, we may choose the three unit complex numbers $z_i=e^{2i\theta_i}$ as follows:
\begin{align}
 h&=4q_1^2q_2^2-(q_3^2-q_1^2-q_2^2)^2,\nonumber\\
 z_1&=1,\nonumber\\
 z_2&=\frac{q_3^2-q_1^2-q_2^2+i\sqrt h}{2q_1q_2},\nonumber\\
 z_3&=-\frac{q_1+q_2z_2}{q_3}.
 \label{eq:trianglecomplex}
\end{align}
The triangle inequalities ensure that $h\ge0$, that $|z_i|=1$, and that $\sum_iq_i z_i=0$. The discrete signs that were absorbed in the initial rephasing are restored by multiplying the appropriate $z_i$ by $\pm1$. The quantities $u_i=\Rea z_i$ are rational functions of the input coefficients, while $\Ima z_i$ consists of a rational factor multiplying $\sqrt h$.

The physical doublet phases $\theta_i$ are half of the arguments of $z_i$. We choose $\theta_i\in[-\pi/2,\pi/2]$, with $\theta_i=\pi/2$ when $z_i=-1$, and introduce $\sigma_i=\sgn(\sin\theta_i)$. Then
\begin{align}
 \cos\theta_i&=\sqrt{(1+u_i)/2},\nonumber\\
 \sin\theta_i&=\sigma_i\sqrt{(1-u_i)/2}.
 \label{eq:halfphases}
\end{align}
If $h>0$, the sign $\sigma_i$ follows from the imaginary part of $z_i$; if $h=0$, it is fixed by the endpoint convention above. The cosine of a phase difference therefore has the same sign as
\begin{align}
 &\sqrt{(1+u_i)(1+u_j)}\nonumber\\
 &\hspace{8mm}+\sigma_i\sigma_j\sqrt{(1-u_i)(1-u_j)},
 \label{eq:cossign}
\end{align}
and the corresponding sine has the sign of
\begin{align}
 &\sigma_i\sqrt{(1-u_i)(1+u_j)}\nonumber\\
 &\hspace{8mm}-\sigma_j\sqrt{(1+u_i)(1-u_j)}.
 \label{eq:sinsign}
\end{align}
The sign of an expression $a\sqrt{x}+b\sqrt{y}$, with $a,b\in\{-1,0,1\}$ and nonnegative rational $x,y$, can be decided without evaluating radicals. When the coefficients have the same nonzero sign, the answer is immediate; when they have opposite signs, the comparison reduces to $x-y$, with zero terms treated separately. Thus all phase-sign decisions reduce to exact rational-function comparisons. The signs of the charged overlaps additionally include the signs of $\nu/\alpha_{ij}$ or $\nu/\beta_{ij}$.

Finally, the squared real and imaginary overlaps are nonnegative rational functions with finite rational limits. Taking their nonnegative square roots and restoring the exact signs yields their limiting values even if separate factors are singular. The construction also works when a nondegenerate triangle approaches a collinear configuration. It avoids subtracting numerical approximations to radicals whose individual terms may be large or divergent.

\section{Validation and reproducibility}\label{app:validation}

In this appendix we describe the checks performed on the exact boundedness-from-below (BFB) criterion and give the information needed to reproduce them. We wish to stress at the outset that the numerical and algebraic tests have different roles. The exact code determines the minimum by implementing the 28 stationary branches derived in Secs.~III--V; a separate decision checker verifies that the prescribed branch tests were executed correctly. An attaining-field checker confirms that a reported value is realized by a physical Higgs configuration. Neither verification, taken alone, proves that no lower value exists. This conclusion relies on the completeness argument of Sec.~V or, for the analytical examples of Sec.~VII, on an independent lower-bound identity together with a field configuration saturating it. The direct numerical minimization described below provides an additional, independent check of the implementation, not a proof of global optimality.

\subsection{Test potentials and numerical procedure}

We considered 398 sets of the twelve real quartic couplings, $p=(a,b,c,d)$, deliberately chosen to test both ordinary and exceptional parameter regions. Each coefficient was stored as an exact rational number, with the pair ordering fixed to $(12,23,31)$. The input file is \path{validation/validation_scripts/inputs.json}; the generator is \path{validation_scripts/make_validation_inputs.py}, using the recorded random seed \texttt{2026101606}. The tested couplings are not intended as a statistical representation of phenomenologically viable models.

Of these inputs, 96 have random coefficients generated by dividing integers in $[-30,30]$ by independently selected members of $\{5,7,9,11\}$. To probe exceptional cases, 48 inputs have vanishing phase-sensitive coefficients, another 48 have one or more singular angular blocks $c_e=\pm d_e$, and 24 have a singular radial matrix obtained with $a=(2,5,10)$ and $b=(3,5,4)$. We also considered 24 scale-separated inputs, obtained by multiplying selected couplings by powers $10^n$ with $-6\leq n\leq6$. These cases test whether exact arithmetic and the deformation prescription remain reliable when coefficients vanish or differ greatly in magnitude.

Since many random potentials have an immediately unstable single-doublet direction, we included 32 backgrounds chosen to favor charge-breaking minima. They use $a_i\in\{9,\ldots,20\}$, $b_e\in\{-3,\ldots,3\}/10$, $c_e\in\{8,\ldots,16\}/10$, and $d_e=\pm\{5,\ldots,25\}/100$. Three further potentials were constructed from each background by shifting its normalized minimum, as returned by the exact solver, to $-1/50$, $0$, and $1/50$. The resulting 96 cases test consistency under shifts but do not constitute independent determinations of their stability boundaries. Finally, we used three analytically solved families from Sec.~VII, at scales $1$ and $7/3$, and shifted their known minima to $-1/50$, $0$, $1/50$, $-10^{-20}$, and $10^{-20}$. These 30 cases have independently established reference values. Shifting or rescaling a potential does not produce a new independent minimization landscape, a distinction we retain in interpreting the results.

For each input we minimized the quartic directly over the twelve real components of the three Higgs doublets. The function and its analytic gradient were implemented separately from the 28-branch solver in \path{validation_scripts/field_objective.py}. We used 40 independently seeded, unit-normalized random starting points and SciPy's BFGS algorithm, with \texttt{gtol=1e-10} and at most 1500 iterations. All termination flags, final values, gradients, and selected configurations were recorded. We compare the smallest numerical value, $V_{\rm best}$, with the exact-algorithm result, $m$, through
\begin{equation}
\epsilon_{\rm num}=\frac{|V_{\rm best}-m|}{s_p},\qquad s_p=\max\bigl(1,\max_i|p_i|\bigr).
\label{eq:app-validation-error}
\end{equation}
The normalization by $s_p$ accounts for the different overall magnitudes of the couplings. The comparison tolerance $10^{-7}$ was used only for the numerical check and does not enter the exact sign classification.

\subsection{Results of the numerical tests}

Table~\ref{tab:app-validation} summarizes the 398 input potentials and their exact-algorithm classifications. A negative $m$ indicates an unstable quartic direction, $m=0$ a nonnegative quartic with a flat direction, and $m>0$ a strictly positive quartic. The large number of negative cases follows from the deliberately broad sampling and should not be interpreted as a prediction for the physically relevant parameter space.

\begin{table}[t]
\caption{Exact-algorithm classifications of the test potentials. The classifications for the shifted analytical examples additionally have independent sharp reference values. The cases where $m \leq 0$ correspond to non-physical potentials.}\label{tab:app-validation}
\begin{ruledtabular}
\begin{tabular}{lrrrr}
Input class & $N$ & $m<0$ & $m=0$ & $m>0$\\
Generic rational & 96 & 94 & 1 & 1\\
Vanishing phase coefficients & 48 & 48 & 0 & 0\\
Singular angular blocks & 48 & 47 & 0 & 1\\
Singular radial matrix & 24 & 2 & 0 & 22\\
Scale-separated couplings & 24 & 23 & 0 & 1\\
Analytical-reference shifts & 30 & 12 & 6 & 12\\
Charge-breaking backgrounds & 32 & 0 & 0 & 32\\
Solver-centered shifts & 96 & 32 & 32 & 32\\
\hline
Total & 398 & 258 & 39 & 101
\end{tabular}
\end{ruledtabular}
\end{table}

The two exact evaluation routes agree on all 398 inputs. Direct field-space minimization also agrees with the reported exact values within the specified normalized tolerance. The largest normalized discrepancy was $3.7240\times10^{-13}$. For scale-separated couplings the largest absolute discrepancy reached $1.0157\times10^{-6}$, whereas outside that class it was approximately $2.11\times10^{-15}$. We therefore quote agreement relative to the coupling scale rather than a universal absolute precision. The 398 inputs correspond to 15,920 numerical starting points; 9,144 BFGS runs returned non-success termination flags, although they all provided finite recorded values. Repeating a local search from many initial points increases its usefulness as a consistency check but does not turn it into a certified global minimization.

The charge-breaking sample is especially relevant to our analysis. Of its 32 backgrounds, the reported minimizing branches comprise 20 full-support charged-collinear, four charged-triangle, and eight split configurations. These examples were selected to probe structures that a neutral-only analysis would miss. For the analytically solved examples, the exact lower bounds and attaining fields of Sec.~VII furnish an independent check of the minimum. We also construct additional test potentials by shifting the original potential by a constant multiple of \(S^2\). Suppose our algorithm predicts a minimum \(\widehat m\), and we wish to shift this minimum to a chosen value \(\delta\). We define
\[
\widetilde V=V+(\delta-\widehat m)S^2.
\]
Since \(S=1\) for normalized fields, the true minimum of the shifted potential is
\[
\widetilde m=m-\widehat m+\delta,
\]
where \(m\) is the actual global minimum of the original potential. Therefore, the shifted minimum equals \(\delta\) only if our prediction is correct, namely \(\widehat m=m\).

If we use the same algorithm to calculate both \(\widehat m\) and the minimum of the shifted potential, obtaining \(\delta\) confirms that the algorithm responds consistently to the shift. However, it does not independently establish that \(\widehat m\) was the correct global minimum, since the same error could be present in both calculations. We therefore use these shifted potentials to test the internal consistency of our method, while relying on independently established analytical minima for rigorous validation.

\subsection{Exact evaluation and comparison with StableWein}

The exact implementation first attempts the stationary systems at the original couplings. If a denominator or bordered matrix is exactly singular, the calculation is repeated using the infinitesimal deformation described in Sec.~V. All support, phase, and Gram-positivity decisions are made using exact rational arithmetic; no small numerical regulator or threshold on the final minimum is introduced. This procedure is implemented in \path{validation_scripts/exact_dispatch.py}. Among the 398 inputs, 282 were treated directly and 116 required the exact deformation. 

We also compared selected inputs with the public StableWein implementation \cite{StableWein}. In translating conventions, the pair order changes from $(12,23,31)$ to $(23,31,12)$, while allowed sign rephasings place the invariant product sign of the phase coefficients in a common source phase $0$ or $\pi$. The mapping was checked for all 27 sign/zero patterns. Eight selected mode-four calls were executed using Wolfram 15.0.1, of which four were classified by preliminary filters and four reached numerical minimization. All eight sign classifications agreed with our criterion. The individual input identifiers, execution times, and deciding stages are preserved in \path{validation/validation_scripts/stablewein_results.json}; the comparison driver is \path{validation_scripts/run_stablewein_benchmark.wl}. The recorded Wolfram-source checksum and the headless execution adaptations are included in the archived comparison data. Since these are selected runs performed in different environments, they do not establish either a population accuracy or a speed advantage over StableWein. The comparison principally checks convention mapping and consistency on selected charged configurations.

Further checks include a finite-difference comparison of 24 gradient components, with maximum residual $9.42\times10^{-11}$, exact evaluation of five analytical examples, and rejection of five malformed-input classes, including when Python assertions are disabled. These tests are useful for detecting implementation errors but do not independently establish the completeness of the 28 branches.

\subsection{Reproducing the calculations}

The code, exact coupling inputs, output records, and selected certificates are publicly available at \url{https://github.com/fayez0/Physics_Papers-Real_3HDM_BFB}. The archive contains the core implementations in \path{code/}, analytical certificates in \path{certificates/}, illustrative inputs in \path{examples/}, and executable validation scripts in \path{validation_scripts/}. The corresponding recorded inputs and results are in \path{validation/validation_scripts/}. The Python dependencies are listed in \texttt{requirements-tested.txt}.

From the published \path{Code/} directory, the following commands invoke the exact example and audit the archived validation records:
\begin{flushleft}\ttfamily\footnotesize
python validation\_scripts/exact\_dispatch.py\\
\hspace*{5mm}examples/charged\_flat.json\\
python validation\_scripts/audit\_revision.py\\
python validation\_scripts/summarize\_validation.py
\end{flushleft}
The archived inputs are stored in \path{validation/validation_scripts/inputs.json} and the per-start search records in \path{validation/validation_scripts/records.jsonl}. The input generator and numerical searches can be rerun with
\begin{flushleft}\ttfamily\footnotesize
python validation\_scripts/\\
\hspace*{5mm}make\_validation\_inputs.py\\
python validation\_scripts/run\_validation.py\\
\hspace*{5mm}--workers 4
\end{flushleft}
Regenerating data may overwrite the saved records, so a separate working copy is recommended. The input generator uses the exact solver to construct the solver-centered shifts; the separately implemented original-field objective used in the BFGS comparison does not invoke the branch solver.

Finally, we emphasize the scope of the evidence. Exact decision auditing demonstrates faithful execution of a stated criterion, and attaining-field certificates establish the existence of physical configurations at the quoted values. Independently derived lower-bound identities establish exact global minima for the corresponding analytical examples. The general criterion, however, rests on the mathematical completeness argument in Secs.~III--V. The computational checks reported here support that argument but do not replace its independent mathematical review.

\end{document}